# Graphene-based topological insulator with an intrinsic bulk band gap above room temperature

Liangzhi Kou[1*,†], Binghai Yan[2,3, †], Feiming Hu[1], Shu-Chun Wu[2],Tim O. Wehling[1], Claudia Felser[2], Changfeng Chen[4] and Thomas Frauenheim[1]

1. Bremen Center for Computational Materials Science, University of Bremen, Am Falturm 1, 28359 Bremen, Germany
2. Max Planck Institute for Chemical Physics of Solids, Noethnitzer Str. 40, 01187 Dresden, Germany
3. Max Planck Institute for the Physics of Complex Systems, Noethnitzer Str. 38, 01187 Dresden, Germany
4. Department of Physics and Astronomy and High Pressure Science and Engineering Center, University of Nevada, Las Vegas, Nevada, 89154, United States

*Corresponding author: kouliangzhi@gmail.com

† These authors have contributed equally to this work.

Topological insulators (TIs) represent a new quantum state of matter characterized by robust gapless states inside the insulating bulk gap. The metallic edge states of a two-dimensional (2D) TI, known as quantum spin Hall (QSH) effect, are immune to backscattering and carry fully spin-polarized dissipationless currents. However, existing 2D TIs realized in HgTe and InAs/GaSb suffer from small bulk gaps (<10 meV) well below room temperature, thus limiting their application in electronic and spintronic devices. Here, we report a new 2D TI comprising a graphene layer sandwiched between two $Bi_2Se_3$ slabs that exhibits a large intrinsic bulk band gap of 30 to 50 meV, making it viable for room-temperature applications. Distinct from previous strategies for enhancing the intrinsic spin-orbit coupling effect of the graphene lattice, the present graphene-based TI operates on a new mechanism of strong inversion between graphene Dirac bands and $Bi_2Se_3$ conduction bands. Strain engineering leads to effective control and substantial enhancement of the bulk gap. Recently reported synthesis of smooth graphene/$Bi_2Se_3$ interfaces demonstrates feasibility of experimental realization of this new 2D TI structure, which holds great promise for nanoscale device applications.

Recent discovery of topological insulators (TIs) has generated great interest in probing fundamental mechanisms and practical applications of this new class of materials that can exist in three-dimensional (3D) compounds or in two-dimensional (2D) systems[1-2]. The TIs are distinguished from conversional insulators by their unique gapless surface states protected by the time-reversal or crystalline symmetry. Intensive search for TI materials has identified a number of compounds that are 3D TIs, but to date only HgTe and InAs/GaSb quantum wells have been verified to be 2D TIs[3-7]. A critical drawback of the reported 2D TIs is that their small bulk gaps (<10 meV) are well below room temperature, thus limiting their application in electronic and spintronic devices. Recent theoretical studies have proposed some layered materials that are promising 2D TI candidates, such as silicon or germanium[8] and ultrathin Bi (111) films[9]. These layered materials, however, suffer from structural instability and high surface chemical activity, and have yet to be confirmed by experiment.

Perhaps the most famous 2D TI is graphene, which is a so-called Dirac material in which carriers possess quasi-relativistic dispersion described by the massless Dirac Hamiltonian[1,2,10]. Kane and Mele predicted[11] graphene as the first 2D TI with helical edge states, which is also known as the quantum spin Hall (QSH) insulator, where the spin-orbit coupling (SOC) opens an energy gap at the Dirac point. Graphene has since become a prototypical model system for 2D TIs[1,2]; however, its bulk energy gap is too small to make the predicted QSH effect observable under normal conditions (a low temperature of below 0.01 K is required)[12]. Several proposals have been put forward to enhance the SOC effect of graphene, such as a direct deposition of hydrogen or heavy element adatoms[8-9,13-15]. Meanwhile, $Bi_2Se_3$ and $Bi_2Te_3$ are the most extensively studied 3D TI materials to date[16-18]. Recent experiment has reported that smooth interfaces between these 3D TIs and grapehene have been achieved using vapor-phase deposition and molecular beam epitaxy techniques[19-21]. It has been argued that the proximity effect from the 3D TI materials may lead to considerably enhanced SOC in grapehene[22-23]. These recent developments raised interesting issues regarding the proximity effect on the electronic structure of graphene, but these heterostructures comprising the 3D TI and graphene are topologically trivial insulators, and thus are not expected to exhibit the QSH effect.

In the present work, we propose a new strategy to achieve a significantly improved 2D TI state that exhibits a large bulk gap by constructing quantum well structures where a graphene layer is sandwiched between thin slabs of $Bi_2Se_3$. We show that the strong hybridization between graphene and $Bi_2Se_3$ across the interfaces opens a large energy gap (30-50 meV) and produces a 2D TI state that is driven by a new mechanism involving an inversion of the graphene Dirac bands and the conduction bands of $Bi_2Se_3$, which is distinct from the original Kane-Mele scenario. The topological characteristic of the quantum well structure is confirmed by the calculated nontrivial $Z_2$ index and an explicit demonstration of the topological edge states in the system. We show that the topological properties are robust against variations in the interlayer stacking pattern and cladding materials in the quantum well structure. Our findings represent a major step forward in the study of QSH effect in 2D TIs.

First-principles calculations based on density functional theory were carried out using the Vienna Ab-initio Simulation Package.[24] The exchange-correlation interaction of electrons was treated within the generalized gradient approximation (GGA) of the Perdew-Burke-Ernzerhof type[25]. Pseudopotentials generated by the projector augmented wave method were used for atomic potentials. The SOC was included in the second variational step using the scalar-relativistic eigenfunctions as a basis. A cutoff energy of 400 eV was used for the expansion of wave functions and potentials in the plane-wave basis. *k*-point meshes of 5×5×1 were used for the sampling of the Brillouin zone. We used the supercell method by putting a single layer of graphene between two $Bi_2Te_3$ single quintuple-layer (QL) slabs and introducing a vacuum layer of thickness 10 Å between the cells to minimize artificial intercell interactions. Once a full atomic relaxation was performed, one additional step of self-consistent calculation including the SOC was carried out until the total energy converged to within $10^{-5}$ eV. To describe the van der Waals–type interaction between graphene and the TI surface, we employed a semiempirical correction using Grimme's method because the GGA cannot adequately describe the van der Waals interaction.[26]

Our design of the quantum well (QW) structure places a graphene layer sandwiched between two $Bi_2Se_3$ slabs. The structural symmetry between the two cladding slabs counteracts the substrate-induced potential difference, thus removes the Rashaba effects and band splitting[27-28]. $Bi_2Se_3$ has a quintuple layer (QL) structure, in which neighboring QLs are coupled via weak van der Waals interactions. While thick slabs of $Bi_2Se_3$ show metallic topological states on top and bottom surfaces, thin slabs of 1-3 QLs are normal insulators due to strong interactions between the surface states[28-29]. We have chosen single QLs of $Bi_2Se_3$ as a wide-gap cladding material of the QW structure, as shown in Fig. 1a. We used the experimental in-plan lattice constant of 4.138 Å for $Bi_2Se_3$ and then adjusted the lattice constant of graphene accordingly. The lattice mismatch is about 2.9% with the use of a $\sqrt{3}\times\sqrt{3}$ in-plan supercell for graphene. We considered two stacking patterns between graphene and $Bi_2Se_3$ with the surface Se atoms at the hollow center of the carbon hexagon rings (Fig. 1b) or on top of the carbon atoms (Fig. 1c). We find that the QW structure with the hollow stacking pattern has a lower binding energy with a equilibrium binding distance between the graphene layer and the $Bi_2Se_3$ QL layer of 3.24 Å, which is 0.02~0.03 Å smaller than the value for the on-top stacking pattern. Below we will mainly focus on this more stable QW structure, but we also will discuss the robustness of our findings against variations in the stacking pattern. The structural and electronic parameters are summarized in Table 1.

**Table 1** Calculated interlayer distance $d_{T\text{-}G}$, binding energy $E_b$ per unit cell (defined as the difference between total energy of the QW and the sum of separate constituent components) and nontrivial band gap $E_g$. $a_{Hex}$ is the experimental lattice parameter of the unit cell used in the calculations, and $E_\Gamma$ is the band gap of single QL materials and graphene at the Γ point. H-TI/G/TI and T-TI/G/TI denote the hollow and on-top stacking patterns, respectively, and "None" indicates the system is not a TI.)

| Top layer | Free standing | | H-TI/G/TI | | | T-TI/G/TI | | | H-2TI/G/2TI | | |
|---|---|---|---|---|---|---|---|---|---|---|---|
| | $a_{Hex}$(Å) | $E_\Gamma$(eV) | $d_{T\text{-}G}$ (Å) | $E_b$(eV) | $E_g$(eV) | $d_{T\text{-}G}$ (Å) | $E_b$(eV) | $E_g$(eV) | $d_{T\text{-}G}$ (Å) | $E_b$(eV) | $E_g$(eV) |
| $Bi_2Se_3$ | 4.138 | 0.82 | 3.244 | -0.112 | 0.030 | 3.267 | -0.108 | 0.035 | 3.258 | -0.367 | 0.011 |
| $Sb_2Se_3$ | 4.076 | 0.74 | 3.382 | -0.476 | 0.015 | 3.397 | -0.509 | 0.002 | 3.408 | -0.401 | 0.017 |
| $Bi_2TeSe_2$ | 4.28 | 0.41 | 3.357 | -0.342 | 0.034 | 3.36 | -0.307 | 0.035 | | | |
| $Bi_2Te_2Se$ | 4.218 | 0.47 | 3.386 | -0.387 | None | 3.478 | -0.403 | None | | | |
| Graphene | 4.26 | 0.00 | | | | | | | | | |

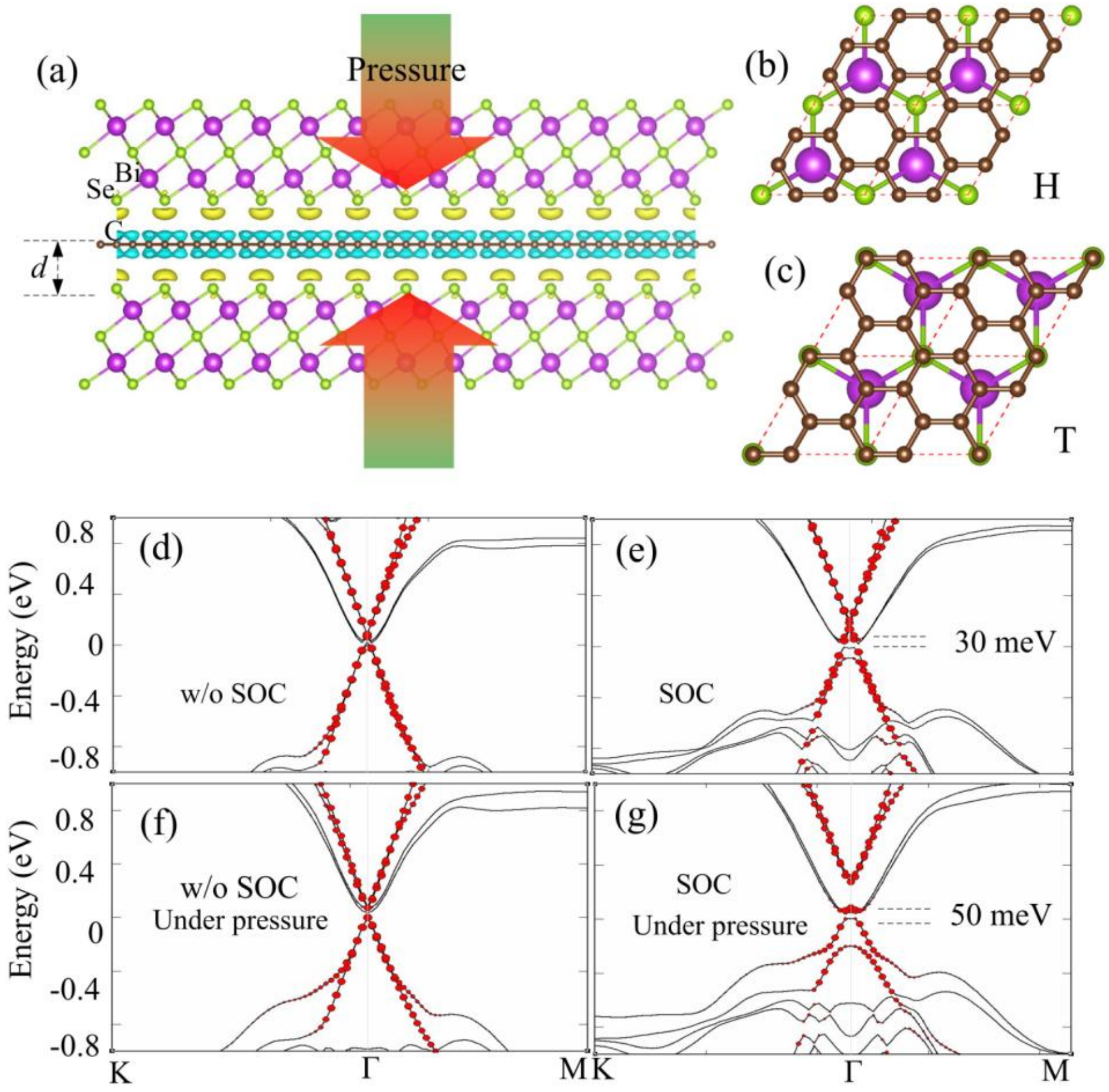

**Figure 1. The proposed QW structure and its electronic bands.** (a) Structural model (side view) of the $Bi_2Se_3$/graphene/$Bi_2Se_3$ quantum well; the blue and yellow isosufaces indicate electron lose and gain [$\Delta\rho=\rho_{Tot}-\rho_{Grap}-\rho_{TI}$ with SOC], respectively, where the isosurface value is $5\times10^{-4}$ e/Å$^3$. (b,c) Top views of the hollow (H) and on-top (T) interlayer stacking patterns. For clarity, we only show the atoms in the graphene and two atomic layers of $Bi_2Se_3$. (d)-(g) Calculated electronic band structures under different physical conditions. The red dots indicate graphene Dirac states. Panels (d, e) show results of the equilibrium structure while panels (f, g) show results at the interlayer distance that has been compressed by 0.45 Å from its equilibrium position.

Without considering the SOC effect, the electronic band structure of the QW is basically a simple superposition of the results for graphene and single QL $Bi_2Se_3$ (Fig. 1d). When the SOC is switched on, the valence bands of $Bi_2Se_3$ are uplifted while the conduction bands are downshifted. The downshifted conduction bands overlap and hybridize with the graphene Dirac states, and the resulting interaction produces a topologically nontrivial inverted band gap of 30 meV (Fig. 1e, see discussions below for a detailed analysis). The strong hybridization is also manifested in the charge transfer from graphene to the $Bi_2Se_3$ QLs, as shown in Fig.1a.

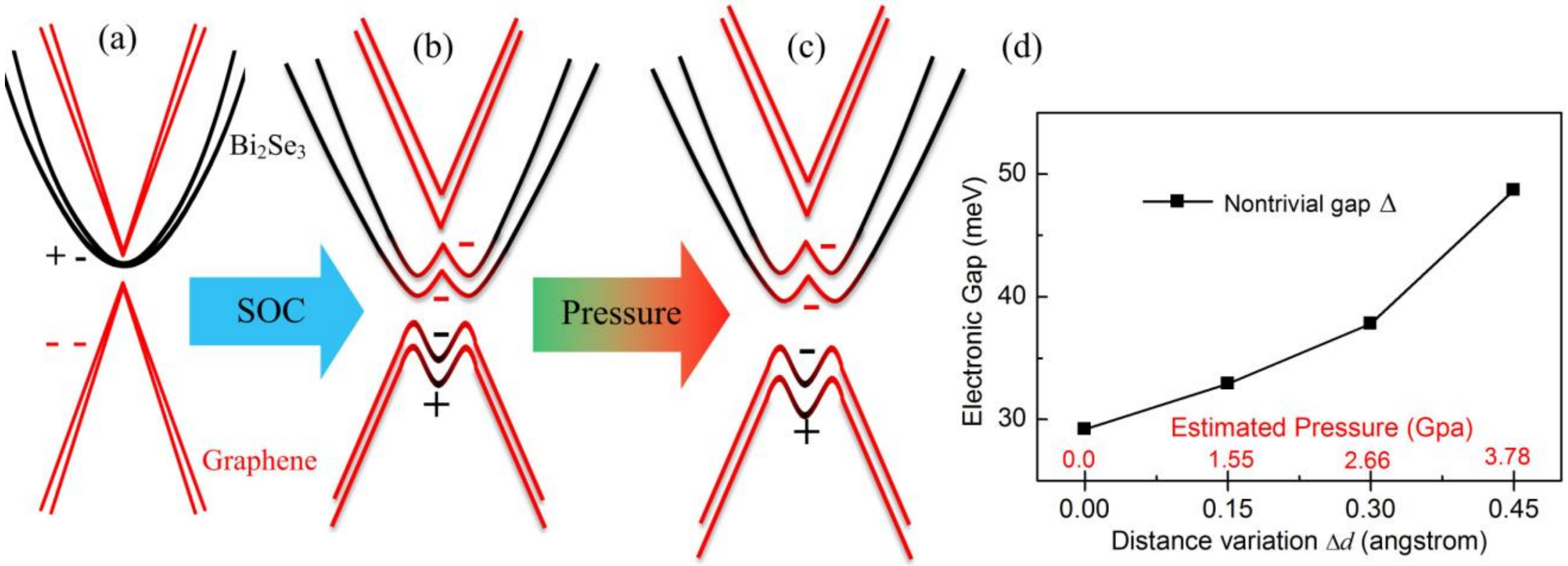

**Figure 2**. **Illustration of the mechanism of the band inversion in the $Bi_2Se_3$/graphene/ $Bi_2Se_3$ QW.** (a) Without considering the SOC effect, the graphene Dirac states open a gap as a consequence of the symmetry breaking in the graphene lattice by the presence of the adjacent $Bi_2Se_3$ QL. The top two valence bands both have "-" parity, and the two $Bi_2Se_3$ conduction bands exhibit one "+" and one "-" parity. (b) With the SOC effect included, the two graphene valence bands are inverted relative to the two $Bi_2Se_3$ conduction bands, resulting in a topologically nontrivial band structure. (c) Under compression, the hybridization between the $Bi_2Se_3$ and graphene states is enhanced, leading to increased nontrivial energy gap. (d) The pressure (or *d* variation) dependence of the energy gap.

To verify the topological nature of our constructed graphene-$Bi_2Se_3$ QW structure, we have calculated the topological $Z_2$ invariant following the parity criteria[30]. With the SOC switched on, the parity eigenvalues were calculated for all the occupied states at time-reversal invariant momenta, i.e., one Γ (0, 0) point and three M (0.5, 0) points. The band inversion at the Γ point induces a nontrivial $Z_2$ = 1, indicating that the graphene-$Bi_2Se_3$ QW is indeed a 2D TI. The underlying mechanism for the obtained TI behavior is illustrated by a band inversion process shown in Fig. 2a-2c. In the absence of the SOC effect, the two Dirac valleys of graphene are projected onto the Γ point, and a tiny energy gap opens up with the two of valence bands having "-" parity and two of conduction bands "+" parity (spin degeneracy is neglected here). The hybridization of the conduction bands from the two $Bi_2Se_3$ QLs leads to the formation of a pair of bonding ("+" parity) and anti-bonding ("-" parity) states. When the SOC is turned on, this pair of $Bi_2Se_3$ states shifts downward and gets inverted with the graphene valence bands, resulting in the nontrivial $Z_2$ index.

External compression can push the layers in the QW structure closer with a reduced interlayer distance, which, in turn, enhances the hybridization between $Bi_2Se_3$ and graphene. From the band structures shown in Fig. 1f and 1g, it is seen that the nontrivial band gap is indeed sensitive to the interlayer distance. The nontrivial band gap Δ increases to 50 meV (nearly 600K) at an interlayer distance compression of 0.45 A, corresponding to a nominal pressure of 3.78 GPa (see Supplementary Information for more details). This gap is considerably larger than those in previously realized QSH systems such as HgTe[3-5] and InAs/GaSb[6-7] QWs. On the other hand, increasing thickness of the $Bi_2Se_3$ layers reduces the QW energy gap. When the thickness of $Bi_2Se_3$ reaches 3QLs on each side, the gap closes due to emerging gapless topological surface states. Therefore, 1QL is the optimal thickness for the $Bi_2Se_3$ layer in the 2D TI QW structure. Recently reported synthesis of smooth graphene/$Bi_2Se_3$ interfaces[19-20] indicates that it is feasible to realize the proposed $Bi_2Se_3$/graphene/ $Bi_2Se_3$ QWs with the current experimental techniques.

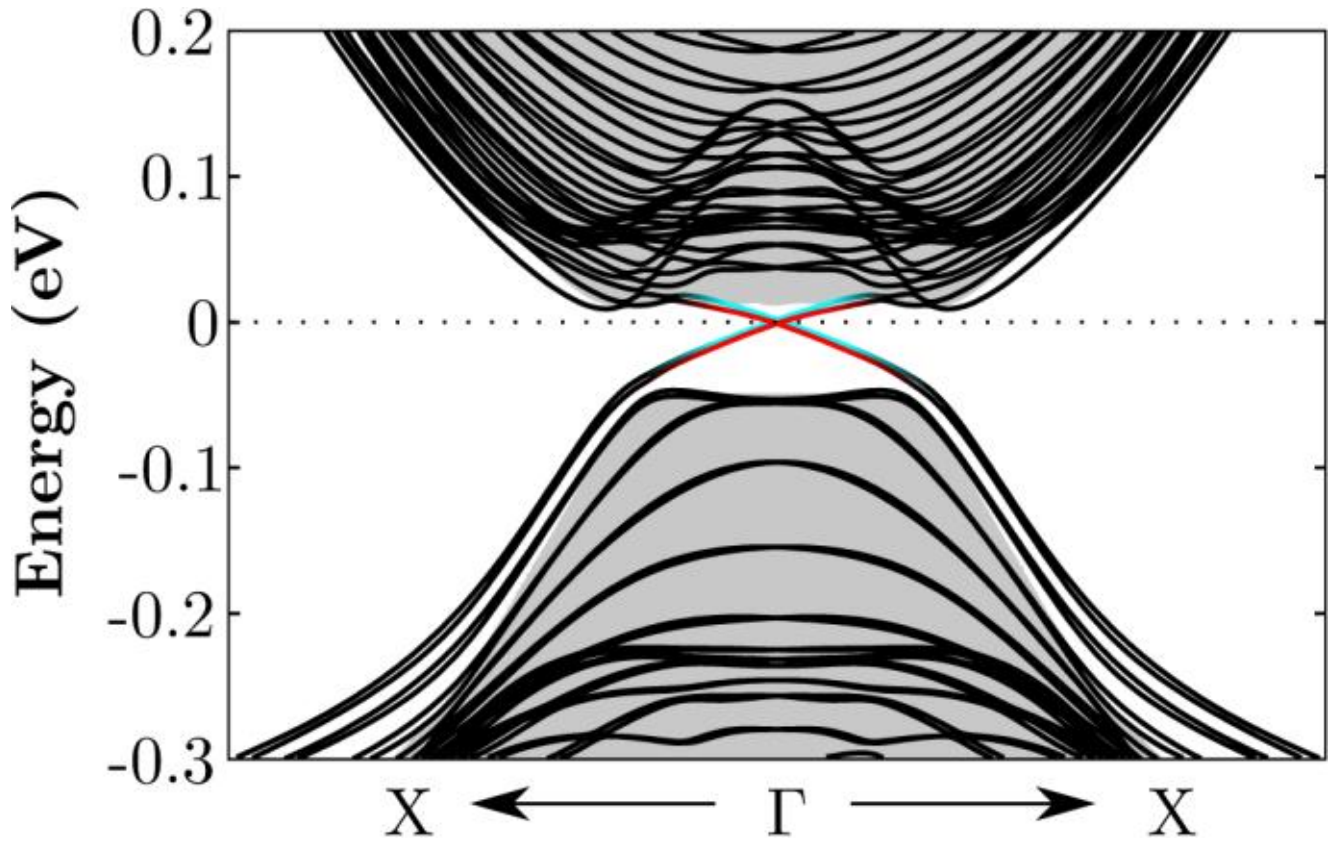

**Figure 3 Topological edge states of the $Bi_2Se_3$/graphene/ $Bi_2Se_3$ QW.** The Dirac helical states denoted by cyan and red solid lines locate at the left and right edges, respectively, on a ribbon model.

A characteristic feature of the QSH state is the existence of 1D helical edge states. We have performed calculations of the edge-state band structure using the Wannier functions extracted from *ab initio* calcualtions[31]. We adopted a ribbon model for the QW structure. The calculated results presented in Fig. 3 show that a pair of gapless edge states is present inside the 2D QW gap at both left and right edges. At a given edge, two counter-propagating edge states exhibit opposite spin-polarizations, a manifestation of the 1D helical nature of the system.

To test the robustness of the topological features of the QW structure against variations of interlayer stacking pattern, we further examined the on-top stacking pattern of the QW structure where the S atoms locate on top of carbon atoms. Our calculations show that the band inversion and parity exchange are well preserved although there are slight variations in the binding energy and interlayer distance (see Table I and Supporting Materials). This insensitivity to variations of interlayer stacking pattern bodes well for experimental synthesis of the proposed 2D TI QW structures.

It should be emphasized that having a TI compound as the QW cladding layer is not a prerequisite for the construction of the new topological insulator. Our proposed graphene-based 2D TI design can be generalized to systems using a topologically trivial insulator like $Sb_2Se_3$ as the cladding layers. Compared to $Bi_2Se_3$, $Sb_2Se_3$ shows a smaller energy gap (12 meV at the equilibrium position) because of the weaker SOC effect of Sb compared to that of Bi. Interestingly, the resulting QW gap is insensitive to the cladding layer thickness and, unlike in graphene-$Bi_2Se_3$ system, the gap persists when the layer thickness increases (Supplementary Information). This is because the $Sb_2Se_3$ slab is always a gapped insulator without topological edge states. In this sense it can be easier to fabricate graphene-$Sb_2Se_3$ QW since it does not impose a stringent requirement on the layer thickness in contrast to those involving $Bi_2Se_3$. A crucial aspect in the design of the topological band inversion in the QW structure is to match the conduction band minima of the cladding layer with the graphene Dirac bands, as shown in Fig. 2. Graphene has a work function (WF) of 4.5~4.8 eV[32] that is close to the values for $Bi_2Se_3$ (4.3 eV) [33] or $Sb_2Se_3$ (4.21 eV) [34]; this makes it easy to match them together. Following this recipe, $Bi_2TeSe_2$ (WF: 4.5 eV) [35] is also a good candidate to fabricate graphene-based QWs (see Table I and Supplementary Information; the nontrivial gap is 35 meV due to enhanced SOC compared to $Bi_2Se_3$). This scenario is further confirmed by the fact that we did not observe such a band inversion in $Bi_2Te_3$/graphene/$Bi_2Te_3$ (WF: 5.3 eV for $Bi_2Te_3$)[36] or $Sb_2Te_3$/graphene/$Sb_2Te_3$ (WF: 5.016 eV for $Sb_2Te_3$)[37] QWs due to the large work function mismatch.

The predicted topological edge states can be verified by transport measurements. A back-gate voltage can tune the Fermi energy of the thin QW structures, and when the Fermi energy lies inside the bulk gap, the appearance of a well-quantized conductance is a clear signature of a perfect edge conductance, similar to that observed in HgTe[4] and InAs/GaSb[38] QWs. In addition, scanning tunneling microscopy may also reveal the peak in the density of states at the QW edge.

In summary, we have identified by ab initio calculations a new family of quantum well structures exhibiting 2D TI phase with a large band gap that far exceeds the gap of current experimentally realized 2D TI materials. We further reveal a new mechanism that involves a strong band inversion of graphene Dirac cone states and $Bi_2Se_3$ conduction bands, which is distinct from the original Kane-Mele scenario. The newly discovered topological phase remains robust against variations of cladding materials and interlayer stacking patterns. These results represent a significant advance in graphene and TI study, and they are expected to stimulate further work to synthesize, characterize, and utilize these new 2D TIs for fundamental exploration and practical application.

**Acknowledgements**

Computation was carried out at HLRN Berlin/Hannover (Germany). L.K. acknowledges financial support by the Alexander von Humboldt Foundation of Germany. B.Y. and C.F. acknowledge financial support from the European Research Council Advanced Grant (ERC 291472). C.C. was supported by the Department of Energy Cooperative Agreement DE-NA0001982.

**Additional information**

Supplementary information is available in the online version of this paper.